\documentclass[letterpaper,10pt,final]{article}
\usepackage{amssymb, amsmath}
\usepackage{graphicx}

\title{Constraining a matter-dominated cosmological model with bulk viscosity
proportional to the Hubble parameter}
\author{Arturo Avelino and Ulises Nucamendi}
\date{\emph{\small Instituto de F\'{\i}sica y Matem\'aticas\\
Universidad Michoacana de San Nicol\'as de Hidalgo \\
Edificio C-3, Ciudad Universitaria, CP. 58040\\
Morelia, Michoac\'an, M\'exico}}

\begin{document}

\maketitle

\begin{abstract}
We present and constrain a cosmological model where the only
component is a pressureless fluid with bulk viscosity as an
explanation for the present accelerated expansion of the universe.
We study the particular model of a bulk viscosity coefficient
proportional to the Hubble parameter: $\zeta_{\rm m}=\zeta H$, where
$\zeta=$ constant. The possible values of $\zeta$ are constrained
using the SNe Ia Gold 2006 sample, the CMB shift parameter $R$ from
the three-year WMAP observations, the Baryon Acoustic Oscillation
(BAO) peak $A$ from the Sloan Digital Sky Survey (SDSS) and the
Second Law of Thermodynamics (SLT). It was found that this model is
in agreement with the SLT using only the SNe Ia test. However when
the model is constrained using the three cosmological tests together
(SNe+CMB+BAO) the results are: 1.- The model violates the SLT, 2.-
It predicts a value of $H_0 \approx 53 \; {\rm km \cdot sec^{-1}
\cdot Mpc^{-1}}$ for the Hubble constant, and 3.- We obtain a bad
fit to data with a $\chi^2_{{\rm min}} \approx 532 \;$
($\chi^2_{{\rm d.o.f.}} \approx 2.92$). These results indicate that
this model is viable just if the bulk viscosity is triggered in
recent times.
\end{abstract}

We present a flat cosmological model which component is a
pressureless fluid made of baryon and dark matter components with a
bulk viscosity coefficient $\zeta_{\rm m}$ proportional to the
Hubble parameter, i.e.,  $\zeta_{\rm m}=\zeta H, \,$ where $\zeta=$
constant and $H$ is the Hubble parameter, to explain the present
accelerated expansion of the universe. The subscript ``m'' stands
for \emph{matter} with bulk viscosity. A bulk viscosity coefficient
can produce a positive term in the second Friedmann equation that
induces an acceleration \cite{Misner1973,Fabris2006} (i.e.,
$\ddot{a} \geq 0 \;$, with $a$ denoting the scale factor and the
dots derivatives with respect to the cosmic time). Similar models or
analysis have been proposed also in
\cite{Fabris2006,ArturoUlisesProc2008,
ArturoUlises2008,Colistete2007,XinHe2006a}.

    \paragraph{A bulk viscous matter-dominated universe}

The energy-momentum tensor of a fluid composed by \emph{only} matter
(hereinafter we call \emph{matter} to baryon and dark matter
components together) with bulk viscosity $\zeta_{{\rm m}}$ is
defined by $T_{\mu\nu}=\rho_{{\rm m}} u_\mu u_\nu +
(g_{\mu\nu}+u_\mu u_\nu)P^*_{{\rm m}}$, where $P^*_{{\rm m}} \equiv
P_{{\rm m}}- 3\zeta_{{\rm  m}} \left( \dot{a}/a \right)$ (see e.g.
\cite{Misner1973}). Here $P_{{\rm m}}$ is the pressure of the matter
fluid and the function $a$ the scale factor.  The four-velocity
vector $u_\mu$ is that of a comoving observer that measures the
pressure $P^*_{{\rm m}}$ and the density $\rho_{{\rm m}}$ of the
matter fluid. Assuming pressureless matter $P_{{\rm m}}=0$ implies
$P^*_{{\rm m}} =-3\zeta_{{\rm  m}} \left( \dot{a}/a \right)$.

On the other hand, the conservation of energy has the form $ a ( d
\rho_{{\rm m}}/da) = 3\left( 3\zeta_{{\rm m}} H -\rho_{{\rm m}}
\right)$, where $H \equiv \dot{a}/a \;$ is the Hubble parameter.
Using the ansatz $\zeta_{{\rm m}}=\zeta H$ for the parametrization
of the bulk viscosity\footnote{The explicit form of the bulk
viscosity is assumed a prior. See for instance
\cite{Fabris2006,ArturoUlisesProc2008,Colistete2007,XinHe2006a}.}
and  the first Friedmann equation $H^2= \left(8\pi G/3
\right)\rho_{\rm m}\,$ (assuming a spatially flat geometry for the
universe), the conservation equation becomes $a (d \rho_{\rm
m}/da)=3\rho_{\rm m} \left[ \left(8\pi G/3 \right)\zeta-1\right]$.
The solution in terms of the redshift\footnote{The superscript ``0''
(zero) represents a quantity evaluated today and the relation
between the redshift $z$ and the scale factor $a$ is $(1+z)=1/a$.}
`$z$' is $\rho_{\rm m}(z)=\rho^0_{\rm m} (1+z)^{3-\tilde{\zeta}}$,
where we have defined the dimensionless bulk viscous coefficient
$\tilde{\zeta} \equiv 8 \pi G \zeta$ and $\rho^0_{\rm m}$ is the
value of the density of matter evaluated today. We substitute this
solution into the first Friedmann equation so it becomes $H^2=
\left(8\pi G/3 \right)\rho^0_{\rm m} (1+z)^{3-\tilde{\zeta}}$.
Dividing the last expression by the \emph{critical density} today
$\rho^0_{\rm{crit}}\equiv 3H^2_0/8\pi G$ we obtain
\begin{equation}
H(z)= H_0 (1+z)^{\frac{1}{2}(3-\tilde{\zeta})}
\end{equation}
\noindent where $H_0$ is the Hubble constant. In this model the
matter density parameter is $\Omega^0_{\rm m} \equiv \rho^0_{\rm
m}/\rho^0_{\rm{crit}}=1$ because the matter is the only component of
the universe.

        \paragraph{Cosmological tests}

We constrain the possible values of the bulk viscosity
$\tilde{\zeta}$ using the following  cosmological tests: the Gold
2006 SNe Ia data sample \cite{Riess2006} composed by 182 SNe Ia, the
Cosmic Microwave Background (CMB) shift parameter `$R$' from the
three-year WMAP observations \cite{WangMukherjee2006}, the Baryon
Acoustic Oscillation (BAO) peak `$A$' from the Sloan Digital Sky
Survey (SDSS) \cite{Eisenstein2005} and the Second Law of
Thermodynamics (SLT) \cite{Misner1973}.

For the SNe Ia test it is defined the observational luminosity
distance \cite{TurnerRiess2002,Riess2004}  in a flat cosmology as
$d_L = c(1+z)H^{-1}_0 \int_0^z E(z')^{-1} \; dz'$, where  $E(z)
\equiv H(z)/H_0$ and $c$ the speed of light. The \emph{theoretical
distance moduli} for the $i$-th supernovae with redshift $z_i$ is $
\mu(z_i)=5\log_{10} [d_L(z_i)/ {\rm Mpc}] +25 \;$. The  statistical
function  $\chi^2_{{\rm SNe}}$ becomes $\chi^2_{{\rm SNe}}
(\tilde{\zeta}, H_0)  \equiv \sum_{k = 1}^{182}
   \left[\mu (z_k , \tilde{\zeta}, H_0) - \mu_k \right]^2 / \sigma_k^2$,
where $\mu_k$ is the \emph{observed} distance moduli for the $k$-th
supernovae and $\sigma_k^2$ is the variance of the measurement.

The CMB shift parameter $R$  is defined as $R \equiv
\sqrt{\Omega^0_{{\rm m}}} \int^{Z_{{\rm CMB}}}_0 E(z')^{-1}\,dz'$,
where $Z_{{\rm CMB}}=1089$ is the redshift of recombination
\cite{WangMukherjee2006,Melchiorri2003}. The observed value of the
shift parameter $R$ is reported to be $R_{{\rm obs}}=1.70 \pm 0.03$
\cite{WangMukherjee2006}.

From the SDSS data, we use the baryon acoustic oscillation (BAO)
peak $A$ defined as $A \equiv \sqrt{\Omega^0_{{\rm m}}}
E(z_1)^{-1/3} \left[ z^{-1}_1
 \int^{z_{1}}_0 E(z')^{-1}\, dz' \right]^{2/3}$,
where $z_1=0.35$ \cite{Eisenstein2005}. The observed value of
SDSS-BAO peak is $A_{{\rm obs}}=0.469 \pm 0.017$
\cite{Eisenstein2005}.

We construct the total $\chi^2$ function as $\chi^2_{{\rm total}}
\equiv \chi^2_{{\rm SNe}} + \chi^2_{{\rm CMB}} + \chi^2_{{\rm
BAO}}$, where $\chi^2_{{\rm SNe}}$ was defined above, $\chi^2_{{\rm
CMB}} \equiv [(R -R_{{\rm obs}})/\sigma_R]^2 \;$ and $ \;
\chi^2_{{\rm BAO}} \equiv [(A -A_{{\rm obs}})/\sigma_A]^2$.

The law of generation of  \textit{local} entropy in the space--time
is found to be $ T \, \nabla_{\nu} s^{\nu} = \zeta_{\rm m}
\nabla_{\nu} u^{\nu} = 3H\zeta_{\rm m}$ \cite{Misner1973}, where $T$
is the temperature, $\nabla_{\nu} s^{\nu} $ is the rate at which
entropy is being generated in a unit volume. The second law of
thermodynamics can be written as $ T\nabla_{\nu} s^{\nu} \geq 0$
which implies that $3H\zeta_{\rm m} \geq 0$. Since $H$ is positive
in an expanding universe then $\zeta_{\rm m}=\zeta H$ has to be
positive in order to preserve the validity of the second law of
thermodynamics. Thus, in the present model a necessary condition is
$\tilde{\zeta} \geq 0$.

\begin{center}
\begin{figure}
  \includegraphics[width=10cm]{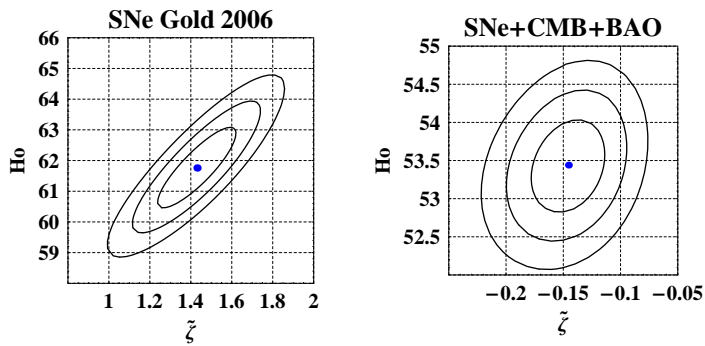}
  \caption{Confidence intervals for the parameters ($\tilde{\zeta}, H_0$)
  for a spatially flat bulk viscous matter-dominated universe. We show the confidence
  intervals of $68.3\% (1\sigma), \; 95.4 \% (3 \sigma)$ and $99.73 \% (5\sigma)$.
  The constraints are derived from the Gold 2006 SNe Ia data sample
  alone (left), and the joint Gold 2006 + CMB + BAO cosmological tests (right).
$H_0$ is in units of ${\rm km} \cdot {\rm sec}^{-1} \cdot {\rm
Mpc^{-1}}$ and $\tilde{\zeta}$ is dimensionless ($\tilde{\zeta}
\equiv 8 \pi G \zeta$).
  }
  \label{ConfInterViscousMatter_z1Ho}
\end{figure}
\end{center}

\begin{center}
\begin{figure}
  \includegraphics[width=6cm]{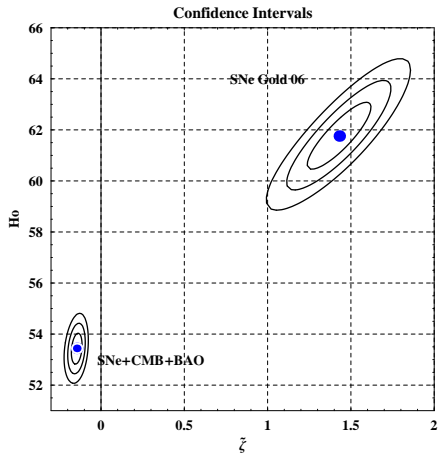}
  \caption{Confidence intervals for the parameters ($\tilde{\zeta}, H_0$)
  for a spatially flat bulk viscous matter-dominated universe. We show the confidence
  intervals of $68.3\% (1\sigma), \; 95.4 \% (3 \sigma)$ and $99.73 \% (5\sigma)$.
  The constraints are derived from the Gold 2006 SNe Ia data sample
  alone and the joint Gold 2006 + CMB + BAO cosmological tests.
  $H_0$ is in units of ${\rm km} \cdot {\rm sec}^{-1} \cdot {\rm
Mpc^{-1}}$ and $\tilde{\zeta}$ is dimensionless ($\tilde{\zeta}
\equiv 8 \pi G \zeta$).}
  \label{ConfInterViscousMatter2_z1Ho}
\end{figure}
\end{center}

      \paragraph{Constraining the bulk viscous cosmological model}
We compute the  \textit{best estimated values} and ``the
goodness-of-fit'' of $\tilde{\zeta}$ and $H_0$ to the data through
$\chi^2$-minimization, using SNe Ia test alone, and also considering
the joint SNe Ia, CMB and BAO tests together. Then we compute the
confidence intervals for $(\tilde{\zeta},H_0)$ to constrain their
possible values. We obtain as best estimates using only the SNe Ia
test: $\tilde{\zeta}=1.43 \pm 0.12$ and $H_0=61.758 \pm 0.86$, with
a $\chi^2_{\rm{min}}=165.135 \; (\chi^2_{\rm{d.o.f.}}= 0.917)$,
where $H_0$ is in units of $\; {\rm km} \cdot {\rm sec}^{-1} \cdot
{\rm Mpc^{-1}}$. On the other hand, using the SNe Ia, CMB and BAO
tests together we obtain $\tilde{\zeta}=-0.1451 \pm 0.02$ and
$H_0=53.425 \pm 0.4$, with a $\chi^2_{\rm{min}}=532.056 \;
(\chi^2_{\rm{d.o.f.}}=2.92)$. It can be seen that when we consider
only the SNe Ia test we obtain $\tilde{\zeta} \geq 0$ with a 99.7\%
confidence level and a reasonable value for the Hubble constant in
the interval $ 58.9 \leq H_0 \leq 64.8$ as well as for
$\chi^2_{\rm{d.o.f.}}=0.917$. However, when the same analysis is
performed using the three cosmological tests together (SNe Ia, CMB
and BAO) we obtain negative values of $\tilde{\zeta}$ with at least
99.7 \% confidence level that disagree with the second law of
thermodynamics, a not so good estimation for $H_0$ of $ 52.1 \leq
H_0 \leq 54.8$, and a bad $\chi^2_{\rm{d.o.f.}}=2.92$. These results
are illustrated in figures \ref{ConfInterViscousMatter_z1Ho} and
\ref{ConfInterViscousMatter2_z1Ho}.

Therefore, we conclude of this model that:
\begin{itemize}
  \item In order to have a viable model for explaining the observed
accelerated expansion of the universe, the bulk viscosity  of the
fluid with the ansatz considered in the present work should be
triggered in recent times ($z \lesssim 2$).
  \item  Since the CMB and BAO tests
  have information for very large redshifts ($z \gg 2$) and due to
  the  fact of that
  this model has a bad fit and violate the second law of thermodynamics
  when we test it using the joint SNe Ia, CMB and BAO tests then
  we conclude that this  is not viable if it is extrapolated for early times
  of the universe as suggested in \cite{Colistete2007}.
\end{itemize}

        \paragraph{Acknowledgments}

     This work is partly supported by grants
     CIC-UMSNH 4.8, PROMEP UMSNH-CA-22, SNI-20733 and COECYT.

        \paragraph*{}

\end{document}